\documentclass{article}
\usepackage{spconf,amsmath,graphicx}
\usepackage{subfigure}
\usepackage{amsfonts}
\usepackage{mathrsfs}
\usepackage{booktabs}
\usepackage{multirow}
\usepackage{caption} 
\usepackage{lipsum}

\title{Multi-Dimensional and Multi-Scale Modeling for Speech Separation Optimized by Discriminative Learning}
\name{Zhaoxi Mu$^{\star}$ \qquad Xinyu Yang$^{\star}$ \qquad Wenjing Zhu$^{\dagger}$}
  
  \address{$^{\star}$ School of Computer Science and Technology, Xi'an Jiaotong University, Xi'an, China \\
      $^{\dagger}$ Du Xiaoman, Beijing, China}

\begin{document}
%
\maketitle
\begin{abstract}
Transformer has shown advanced performance in speech separation, benefiting from its ability to capture global features. However, capturing local features and channel information of audio sequences in speech separation is equally important. In this paper, we present a novel approach named \textbf{I}ntra-\textbf{S}E-\textbf{C}onformer and \textbf{I}nter-\textbf{T}ransformer (ISCIT) for speech separation. Specifically, we design a new network SE-Conformer that can model audio sequences in multiple dimensions and scales, and apply it to the dual-path speech separation framework. Furthermore, we propose Multi-Block Feature Aggregation to improve the separation effect by selectively utilizing information from the intermediate blocks of the separation network. Meanwhile, we propose a speaker similarity discriminative loss to optimize the speech separation model to address the problem of poor performance when speakers have similar voices. Experimental results on the benchmark datasets WSJ0-2mix and WHAM! show that ISCIT can achieve state-of-the-art results.
\end{abstract}
\section{Introduction}

Speech separation is a fundamental task in speech signal processing, aiming to separate sounds mixed by multiple speakers. This study focuses on single-channel speech separation. In previous work, RNN~\cite{luo2020dual,nachmani2020voice}, Transformer~\cite{subakan2021attention} and the combination of both~\cite{chen2020dual,lam2021sandglasset} are used to model speech sequences. However, the inherent sequential characteristic of RNN is not conducive to the parallelization of computation, resulting in low computational efficiency. Transformer is suitable for modeling the global structure of sequences. However, it performs worse than CNN in capturing local information of sequences because CNN can utilize local receptive fields, shared weights, and temporal sub-sampling. 

The audio signal has different levels of context information, such as phonemes, syllables, or words of different granularities. Therefore, we assume that capturing both global and local features of audio sequences is crucial in speech separation. Furthermore, in the time-domain method, the channel (also called feature map) information of the audio signal's feature representation corresponds to the frequency information in the time-frequency domain \cite{luo2019conv,QinZW021}, which is crucial for speech separation. The reason is that audio signals' low and high frequency information have different importance for speech separation \cite{TakahashiM17}. Therefore, capturing the channel information of the audio feature sequence in the separation network is also indispensable.

To achieve these goals, we present SE-Conformer, which combines the strengths of CNN and Transformer, using Conformer~\cite{gulati2020conformer} enhanced by Squeeze-and-Excitation (SE) block~\cite{hu2018squeeze} to model audio sequences. Conformer puts a convolution module after the multi-head self-attention to improve the model's ability to capture local features in the temporal dimension, which has been widely used in speech recognition~\cite{gulati2020conformer}, speech enhancement~\cite{kim2021se,CaoAY22}, and continuous speech separation~\cite{chen2021continuous}. SE block can capture the inter-dependency between channels. This method for simulating channel attention has shown its superiority in capturing channel information of speech sequences~\cite{desplanques2020ecapa}. Overall, SE-Conformer can capture local and global information in the temporal dimension and information in the channel dimension.

Previous studies have shown that features learned from intermediate blocks in the separation network can be employed to enhance the final separation~\cite{nachmani2020voice,byun2021monaural}. Inspired by this, we employ Multi-Block Feature Aggregation (MBFA) to selectively exploit the supplementary information in intermediate blocks by computing the exponentially weighted moving average (EWMA) of the outputs of all blocks.

Furthermore, current speech separation methods do not perform well in separating mixtures of speakers with similar voices. The reason is that the separation network cannot identify the characteristics of each speaker well in this case. Therefore, we use the idea of discriminative learning~\cite{hershey2016deep,fan2019discriminative,fan2020end} to increase the similarity between the estimated and clean sources of the same speaker while decreasing the similarity between the estimated sources of different speakers. Speakers with similar voices can be efficiently distinguished and separated by optimizing the separation network using the proposed loss function.

The contributions of this paper are as follows: (\romannumeral1) We propose a speech separation network composed of SE-Conformer and Transformer, which can effectively model speech sequences in multiple dimensions and scales. (\romannumeral2) We propose Multi-Block Feature Aggregation to improve speech separation by selectively exploiting information from intermediate blocks. (\romannumeral3) We propose speaker similarity discriminative loss that can effectively train models to separate mixtures of speakers with similar voices. (\romannumeral4) Experimental results show that the proposed method achieves state-of-the-art results.

\begin{figure}[t]
    \centering
    \includegraphics[width=0.49\textwidth]{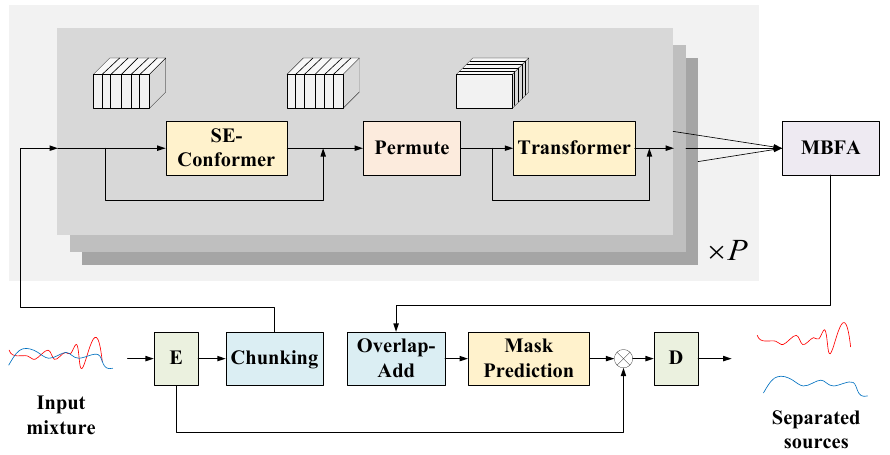}
    \caption{The structure of ISCIT. E and D are the encoder and decoder. Inside the grey box are $P$ Intra-SE-Conformer and Inter-Transformer blocks. $\otimes$ is element-wise multiplication.}
    \label{fig1}
\end{figure}

\section{Methodology}

\subsection{Overall pipeline}

The overall pipeline of ISCIT follows the time-domain dual-path framework \cite{luo2020dual}, as shown in Fig.~\ref{fig1}. The encoder and decoder consist of a $1$-D convolutional layer with $N$ filters and a $1$-D transposed convolutional layer with a symmetric structure. The encoder converts the mixture into the feature representation $z_{\text{e}} \in \mathbb{R}^{N \times T}$ of length $T$. $z_{\text{e}}$ is divided into overlapping chunks $z \in \mathbb{R}^{N\times K\times S}$ with $50\%$ overlap factor in the temporal dimension, where the size and number of chunks are $K$ and $S$. $z$ is fed into $P$ Intra-SE-Conformer and Inter-Transformer blocks consisting of $m$ SE-Conformer layers and $n$ Transformer layers. SE-Conformer and Transformer are employed to process intra-chunk and inter-chunk sequences. The Multi-Block Feature Aggregation (MBFA) module merges the outputs of $P$ blocks. The output is then transformed back to the original sequence by the overlap-add method~\cite{luo2020dual}. A gated convolutional layer with the ReLU activation function is employed to predict the mask. Finally, the decoder converts the masked feature representation into a waveform signal. The following sections describe each component in detail.

\subsection{Intra-SE-Conformer and Inter-Transformer}

\begin{figure}[t]
    \centering
    \includegraphics[width=0.49\textwidth]{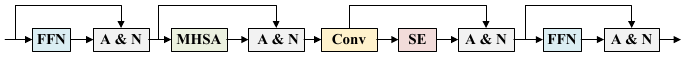}
    \caption{The structure of SE-Conformer. A \& N is Add \& Norm.}
    \label{fig2}
\end{figure}

\textbf{SE-Conformer.} As shown in Fig.~\ref{fig2}, SE-Conformer employs a pair of half-step feed-forward layers (FFN)~\cite{gulati2020conformer} sandwiching a multi-head self-attention module (MHSA), a convolution module (Conv) and an SE block with residual connections. The kernel size of the convolution module of each SE-Conformer block is set to increase successively to extract local information at different scales. The multi-head attention module, convolution module, and SE block are used to capture the global and local information in the temporal dimension and the information in the channel dimension. These three modules are complementary, and their combination can effectively extract information at multiple dimensions and scales in speech sequences.

\textbf{SE block.} Assuming $v$ is the input of the SE block, the output of the SE block is
\begin{equation}
\text{SE}(v)=\sigma(W_{2} f(W_{1} v_{\text{avg}})) \otimes v
\end{equation}
$W_1$ and $W_2$ are the weights of two linear layers. $f$ is the activation function ReLU. $\sigma$ is the sigmoid function. $\otimes$ is element-wise multiplication. SE block first compresses the input $v$ by applying global average pooling on the frame-level dimension to get $v_{\text{avg}}$. The pooling operation can mask the temporal information, allowing the network to efficiently extract the channel information \cite{hu2018squeeze}. Then two linear layers are applied to recalculate the weights of each channel. The output is mapped to $(0, 1)$ through the sigmoid function as the weight of each channel of $v$. These weights strengthen the features of important channels and weaken the non-important through channel-wise multiplication with $v$.

The elements in intra-chunk sequences are continuous to original audio sequences, so both local and global information of intra-chunk sequences are important. As mentioned earlier, SE-Conformer can capture local and global information in the temporal dimension and channel information of sequences, so we use SE-Conformer to build intra-chunk sequences. Since the elements in inter-chunk sequences are non-continuous to original audio sequences, capturing the global dependency of sequences is more important. Therefore, multiple Transformer layers are stacked to model inter-chunk sequences.

\subsection{Multi-Block Feature Aggregation}

Feature aggregation is performed to utilize the information in the intermediate blocks of the separation network to improve performance. After the input passes through $P$ Intra-SE-Conformer and Inter-Transformer blocks, $P$ feature representations can be obtained. Since the feature representation of later blocks in the network is more helpful for separation, we use the exponentially weighted moving average to set the weight of each block to increase sequentially.
\begin{equation}
R_{j}=\beta \times Y_{j}+(1-\beta)\times R_{j-1}, j=1,\ldots,P
\end{equation}
where $R_{j}$ is the aggregated feature representation of the first $j$ blocks and $Y_{j}$ is the output feature representation of the $j$th block. $\beta$ is the weight decay coefficient.

\subsection{Training objective}

The model is trained by maximizing the SI-SNR~\cite{le2019sdr}. Furthermore, a speaker similarity discriminative loss is also used during training. The basic idea is to increase the voice similarity of the same speaker and decrease the voice similarity of different speakers. Specifically, a pre-trained speaker recognition network is used to calculate the speaker embeddings of the clean and estimated sources. The cosine similarity of the speaker embeddings is regarded as the degree of speaker similarity. Without loss of generality, taking the case of two speakers as an example, the sum of the following three dot products is calculated as the loss function:

\begin{equation}
L_{\text{spk}}=-\langle s_1,\hat{s}_1\rangle -\langle s_2,\hat{s}_2\rangle+\langle \hat{s}_1,\hat{s}_2\rangle
\end{equation}
where $s_1$, $\hat{s}_1$, $s_2$ and $\hat{s}_2$ denote clean and estimated source of the two speakers. $\langle \cdot,\cdot \rangle$ denotes dot product operation, which measures the similarity between different speaker embeddings. The first two dot products represent the similarity of the clean and estimated source of the two speakers. The third dot product represents the similarity of the estimated source of the two speakers. The network is optimized by minimizing this discriminative loss function. The total loss function is $L_{\text{SI-SNR}}+\alpha L_{\text{spk}}$, where $\alpha$ is a coefficient that balances the weights between these two loss functions.

\section{Experiments}

\subsection{Experimental setup}

Methods are trained and evaluated on the WSJ0-2mix~\cite{hershey2016deep} and WHAM!~\cite{wichern2019wham} datasets. WHAM! is a noise version of WSJ0-2mix derived from two-speaker mixtures from WSJ0-2mix plus actual ambient noise samples. SI-SNRi and SDRi are reported as objective measures of speech separation.

We implement two models with different numbers of parameters. The model with fewer parameters stacks $4$ layers of SE-Conformer and $6$ layers of Transformer in each Intra-SE-Conformer and Inter-Transformer block denoted as ISCIT, i.e., $m=4$, $n=6$. The model with more parameters stacks $8$ layers of SE-Conformer and Transformer in each Intra-SE-Conformer and Inter-Transformer block denoted as ISCIT (large), i.e., $m=n=8$. By default, we experiment with the model with fewer parameters to reduce computational complexity. 

The kernel size, stride, and the number of filters $N$ of the convolutional layer in the encoder and decoder are set to $16$, $8$, and $256$. The chunk size $K$ is set to $250$. $P$ is set to $3$. In ISCIT, the kernel sizes in the convolutional layers of SE-Conformer are set to $13$, $15$, $17$, and $19$. In ISCIT (large), the kernel sizes are set to $13$, $15$, $17$, $19$, $21$, $23$, $25$, and $27$. The number of heads in the multi-head self-attention is set to $8$. The weight decay coefficient $\beta$ is set to $0.6$. An improved time-delay neural network (TDNN)~\cite{desplanques2020ecapa} trained on the VoxCeleb2 corpus~\cite{chung2018voxceleb2} which contains more than $6000$ speakers is used as the speaker embedding extraction network. The coefficient $\alpha$ of $L_{\text{spk}}$ is set to $1$. Adam optimizer with an initial learning rate of $1.5\text{e}^{-4}$ is used to train the model. After $50$ epochs, the learning rate will be halved if the loss of the validation set does not decrease over three consecutive epochs. Dynamic mixing (DM)~\cite{zeghidour2021wavesplit} is used as data augmentation.

\begin{table}[th]
    \small
    \caption{SI-SNR improvement (dB) on WSJ0-2mix when permuting SE-Conformer (S) and Transformer (T) in intra-block (Intra) and inter-block (Inter).}
    \label{tab5}
    \centering
    \begin{tabular}{lcc}
        \toprule
        \textbf{Method}  & Intra-S & Intra-T\\
        \midrule
        Inter-S & $19.3$           & $18.5$      \\
        Inter-T & $\textbf{21.6}$  & $20.3$      \\
        \bottomrule
    \end{tabular}
\end{table}

\subsection{Results and discussion}

\textbf{The effect of different permutations.} Different permutations of SE-Conformer and Transformer are used as intra-chunk and inter-chunk models to verify the effectiveness of the proposed structure. The results are shown in Table~\ref{tab5}. The permutation we used is shown to have the best performance. Using SE-Conformer to model both intra-chunk and inter-chunk sequences shows worse results. The reason may be that more attention needs to be paid to global information than local information when modeling downsampled inter-chunk sequences. A good result is also achieved using Transformer to model intra-chunk and inter-chunk sequences because the dual-path architecture can partially capture audio sequences' global and local dependencies.

\begin{table}[th]
    \small
    \caption{Ablation analysis of ISCIT on WSJ0-2mix dataset and performance improvement of the proposed method on Sepformer.}
    \label{tab3}
    \centering
    \begin{tabular}{lcc}
        \toprule
        \textbf{Method}  & \textbf{SI-SNRi} & \textbf{SDRi}\\
        \midrule
        ISCIT               & $\textbf{21.6}$  & $\textbf{21.7}$      \\
        w/o SE block          & $21.3$  & $21.4$      \\
        w/o MBFA        & $21.0$  & $21.1$      \\
        w/o $L_{\text{spk}}$     & $21.2$  & $21.3$      \\
         \midrule
        Sepformer~\cite{subakan2021attention} & $20.4$  & $20.5$      \\
        w/ MBFA & $20.9$  & $21.0$      \\
        w/ $L_{\text{spk}}$ & $20.8$  & $20.9$      \\
        \bottomrule
    \end{tabular}
\end{table}

\textbf{Ablation experiments.} Ablation experiments are performed on the SE block, MBFA, and speaker discriminative loss to elucidate the contribution of each component. The results on the WSJ0-2mix dataset are shown in Table~\ref{tab3}. It can be seen that all components contribute to improving the performance of the proposed method. The performance improvement resulting from using MBFA and speaker loss in Sepformer validates the generality of the proposed method. 

\begin{table}[th]
    \small
    \caption{Results of different feature fusion strategies on the WSJ0-2mix dataset.}
    \label{tab6}
    \centering
    \begin{tabular}{lcc}
        \toprule
        \textbf{Fusion strategy}  & \textbf{SI-SNRi} & \textbf{SDRi}\\
        \midrule
        MBFA               & $\textbf{21.6}$  & $\textbf{21.7}$      \\
        Summation          & $21.4$  & $21.5$      \\
        Concatenation       & $21.3$  & $21.4$      \\
        \bottomrule
    \end{tabular}
\end{table}

\textbf{Comparison of different feature fusion strategies.} We also compare MBFA with two other feature fusion strategies: simple summation and concatenation, as shown in Table \ref{tab6}. The results show that the exponentially weighted moving average can more selectively fuse the output feature representations of multiple blocks, resulting in better performance than the other two strategies.

\begin{table}[th]
    \small
    \caption{SI-SNR improvement (dB) for different speaker gender combinations on the WSJ0-2mix dataset. FF, MM, FM, and AVG represent female-female, male-male, female-male, and average.}
    \label{tab4}
    \centering
    \begin{tabular}{lcccc}
        \toprule
        \textbf{Method} & \textbf{FF} & \textbf{MM} & \textbf{FM} & \textbf{AVG}\\
        \midrule
        ISCIT & $\textbf{19.6}$ & $\textbf{20.9}$ & $\textbf{22.0}$ & $\textbf{21.6}$  \\
        w/o $L_{\text{spk}}$ & $19.2$ & $20.5$ & $21.8$ & $21.2$ \\
        \midrule
        Conv-TasNet~\cite{luo2019conv} & $13.2$ & $14.7$ & $15.7$ & $15.3$ \\
        w/ $L_{\text{spk}}$ & $\textbf{13.7}$ & $\textbf{15.5}$ & $\textbf{16.1}$ & $\textbf{16.5}$  \\
        \midrule
        DPRNN~\cite{luo2020dual} & $17.9$ & $18.5$ & $19.2$ & $18.8$ \\
        w/ $L_{\text{spk}}$ & $\textbf{18.4}$ & $\textbf{19.2}$ & $\textbf{19.7}$ & $\textbf{19.3}$  \\        
        \bottomrule
    \end{tabular}
\end{table}

\textbf{Results on different gender combinations.} We separately conduct experiments on the mixture of speakers with different gender combinations to understand how speaker discriminative loss contributes to speech separation models. The results on the WSJ0-2mix dataset are shown in Table~\ref{tab4}. It can be seen that speaker discriminative loss improves the mixing of same-sex speakers more obviously for all speech separation models. This indicates that the speaker loss can effectively deal with the situation where the speakers have similar voices by reducing the similarity of the speaker embeddings. This loss function does not contribute much in the FM case since the similarity of embedding vectors between speakers of different genders is already very low. Meanwhile, the performance improvements on multiple models demonstrate the generality of the proposed speaker loss again.

\begin{table}[th]
  \small
  \caption{Model size, SI-SNR and SDR improvements (dB) on the WSJ0-2Mix dataset.}
  \label{tab1}
  \centering
  \begin{tabular}{lrcc}
    \toprule
    \textbf{Method}  & \textbf{\# Params} & \textbf{SI-SNRi} & \textbf{SDRi}\\
    \midrule
    Sandglasset~\cite{lam2021sandglasset}   &  $2.3$M & $20.8$  & $21.0$      \\
    Wavesplit~\cite{zeghidour2021wavesplit}   &  $29$M & $21.0$  & $21.2$ \\
    Wavesplit + DM~\cite{zeghidour2021wavesplit}   &  $29$M & $22.2$  & $22.3$ \\
    Sepformer~\cite{subakan2021attention}   &  $26$M & $20.4$  & $20.5$      \\
    Sepformer + DM~\cite{subakan2021attention}   &  $26$M & $22.3$  & $22.4$      \\
    TFPSNet~\cite{YangLW22} &  $2.7$M & $21.1$  & $21.3$      \\
    SFSRNet~\cite{RixenR22}   &  $59$M & $22.0$  & $22.1$      \\
    SFSRNet + DM~\cite{RixenR22}   &  $59$M & $24.0$  & $24.1$      \\
    \midrule
    ISCIT  & $34.2$M  & $21.6$  & $21.7$      \\
    ISCIT + DM  & $34.2$M  & $23.4$  & $23.5$      \\
    ISCIT (large)  & $58.4$M  & $22.4$  & $22.5$      \\
    ISCIT (large) + DM & $58.4$M  & $\textbf{24.3}$  & $\textbf{24.4}$      \\
    \bottomrule
  \end{tabular}
\end{table}

\textbf{Comparison with other baseline models.} Our method is compared with other well-performing methods in model size and performance on the WSJ0-2mix dataset to verify the effectiveness reasonably. The results are reported in Table~\ref{tab1}. It can be seen that the proposed method ISCIT (large) + DM outperforms all baseline methods. Our method ISCIT (large) + DM also achieves state-of-the-art results when the mixture contains noise, as reported in Table~\ref{tab2}, which validates the generality and robustness of our method.

\begin{table}[th]
  \small
  \caption{SI-SNR and SDR improvements (dB) on the WHAM! dataset.}
  \label{tab2}
  \centering
  \begin{tabular}{lcc}
    \toprule
    \textbf{Method} & \textbf{SI-SNRi} & \textbf{SDRi} \\
    \midrule
    Wavesplit~\cite{zeghidour2021wavesplit} & $15.4$  & $15.8$     \\
    Wavesplit + DM~\cite{zeghidour2021wavesplit} & $16.0$  & $16.5$     \\
    Sepformer~\cite{subakan2021attention} & $15.5$  & $15.8$      \\
    Sepformer + DM~\cite{subakan2021attention} & $16.4$  & $16.7$      \\
    \midrule
    ISCIT  & $16.0$  & $16.4$      \\
    ISCIT + DM  & $16.6$  & $17.0$      \\
    ISCIT (large)  & $16.4$  & $16.8$      \\
    ISCIT (large) + DM  & $\textbf{16.9}$  & $\textbf{17.2}$      \\
    \bottomrule
  \end{tabular}
\end{table}

\section{Conclusion}

In this paper, we propose SE-Conformer for multi-dimensional and multi-scale modeling of audio sequences. Meanwhile, Multi-Block Feature Aggregation is applied to selectively exploit the information of intermediate blocks in the separation network. We also propose speaker similarity discriminative loss to effectively train speech separation networks to separate speakers with similar voices. The experimental results verify the effectiveness of our method. In addition, the proposed Multi-Block Feature Aggregation and speaker similarity discriminative loss are general and can also be used to enhance other speech separation methods.

\vfill\pagebreak

\bibliographystyle{IEEEbib}
\bibliography{refs}

\begin{thebibliography}{10}

\bibitem{luo2020dual}
Yi~Luo, Zhuo Chen, and Takuya Yoshioka,
\newblock ``{Dual-Path RNN: Efficient Long Sequence Modeling for Time-Domain
  Single-Channel Speech Separation},''
\newblock in {\em {ICASSP}}. 2020, pp. 46--50, {IEEE}.

\bibitem{nachmani2020voice}
Eliya Nachmani, Yossi Adi, and Lior Wolf,
\newblock ``{Voice Separation with an Unknown Number of Multiple Speakers},''
\newblock in {\em {ICML}}. 2020, vol. 119 of {\em {Proceedings of Machine
  Learning Research}}, pp. 7164--7175, {PMLR}.

\bibitem{subakan2021attention}
Cem Subakan, Mirco Ravanelli, Samuele Cornell, Mirko Bronzi, and Jianyuan
  Zhong,
\newblock ``{Attention is All You Need in Speech Separation},''
\newblock in {\em {ICASSP}}. 2021, pp. 21--25, {IEEE}.

\bibitem{chen2020dual}
Jingjing Chen, Qirong Mao, and Dong Liu,
\newblock ``{Dual-Path Transformer Network: Direct Context-Aware Modeling for
  End-to-End Monaural Speech Separation},''
\newblock in {\em {INTERSPEECH}}. 2020, pp. 2642--2646, {ISCA}.

\bibitem{lam2021sandglasset}
Max W.~Y. Lam, Jun Wang, Dan Su, and Dong Yu,
\newblock ``{Sandglasset: A Light Multi-Granularity Self-Attentive Network for
  Time-Domain Speech Separation},''
\newblock in {\em {ICASSP}}. 2021, pp. 5759--5763, {IEEE}.

\bibitem{luo2019conv}
Yi~Luo and Nima Mesgarani,
\newblock ``{Conv-TasNet: Surpassing Ideal Time-Frequency Magnitude Masking for
  Speech Separation},''
\newblock {\em {IEEE ACM Trans. Audio Speech Lang. Process.}}, vol. 27, no. 8,
  pp. 1256--1266, 2019.

\bibitem{QinZW021}
Zequn Qin, Pengyi Zhang, Fei Wu, and Xi~Li,
\newblock ``{FcaNet: Frequency Channel Attention Networks},''
\newblock in {\em {ICCV}}. 2021, pp. 763--772, {IEEE}.

\bibitem{TakahashiM17}
Naoya Takahashi and Yuki Mitsufuji,
\newblock ``{Multi-Scale multi-band densenets for audio source separation},''
\newblock in {\em {WASPAA}}. 2017, pp. 21--25, {IEEE}.

\bibitem{gulati2020conformer}
Anmol Gulati, James Qin, Chung{-}Cheng Chiu, Niki Parmar, Yu~Zhang, Jiahui Yu,
  Wei Han, Shibo Wang, Zhengdong Zhang, Yonghui Wu, and Ruoming Pang,
\newblock ``{Conformer: Convolution-augmented Transformer for Speech
  Recognition},''
\newblock in {\em {INTERSPEECH}}. 2020, pp. 5036--5040, {ISCA}.

\bibitem{hu2018squeeze}
Jie Hu, Li~Shen, and Gang Sun,
\newblock ``{Squeeze-and-Excitation Networks},''
\newblock in {\em {CVPR}}. 2018, pp. 7132--7141, {Computer Vision Foundation /
  IEEE Computer Society}.

\bibitem{kim2021se}
Eesung Kim and Hyeji Seo,
\newblock ``{SE-Conformer: Time-Domain Speech Enhancement Using Conformer},''
\newblock in {\em {INTERSPEECH}}. 2021, pp. 2736--2740, {ISCA}.

\bibitem{CaoAY22}
Ruizhe Cao, Sherif Abdulatif, and Bin Yang,
\newblock ``{CMGAN: Conformer-based Metric GAN for Speech Enhancement},''
\newblock in {\em {INTERSPEECH}}. 2022, pp. 936--940, {ISCA}.

\bibitem{chen2021continuous}
Sanyuan Chen, Yu~Wu, Zhuo Chen, Jian Wu, Jinyu Li, Takuya Yoshioka, Chengyi
  Wang, Shujie Liu, and Ming Zhou,
\newblock ``{Continuous Speech Separation with Conformer},''
\newblock in {\em {ICASSP}}. 2021, pp. 5749--5753, {IEEE}.

\bibitem{desplanques2020ecapa}
Brecht Desplanques, Jenthe Thienpondt, and Kris Demuynck,
\newblock ``{ECAPA-TDNN: Emphasized Channel Attention, Propagation and
  Aggregation in TDNN Based Speaker Verification},''
\newblock in {\em {INTERSPEECH}}. 2020, pp. 3830--3834, {ISCA}.

\bibitem{byun2021monaural}
Jaeuk Byun and Jong~Won Shin,
\newblock ``{Monaural Speech Separation Using Speaker Embedding From
  Preliminary Separation},''
\newblock {\em {IEEE ACM Trans. Audio Speech Lang. Process.}}, vol. 29, pp.
  2753--2763, 2021.

\bibitem{hershey2016deep}
John~R. Hershey, Zhuo Chen, Jonathan~Le Roux, and Shinji Watanabe,
\newblock ``{Deep Clustering: Discriminative Embeddings for Segmentation and
  Separation},''
\newblock in {\em {ICASSP}}. 2016, pp. 31--35, {IEEE}.

\bibitem{fan2019discriminative}
Cunhang Fan, Bin Liu, Jianhua Tao, Jiangyan Yi, and Zhengqi Wen,
\newblock ``{Discriminative Learning for Monaural Speech Separation Using Deep
  Embedding Features},''
\newblock in {\em {INTERSPEECH}}. 2019, pp. 4599--4603, {ISCA}.

\bibitem{fan2020end}
Cunhang Fan, Jianhua Tao, Bin Liu, Jiangyan Yi, Zhengqi Wen, and Xuefei Liu,
\newblock ``{End-to-End Post-Filter for Speech Separation With Deep Attention
  Fusion Features},''
\newblock {\em {IEEE ACM Trans. Audio Speech Lang. Process.}}, vol. 28, pp.
  1303--1314, 2020.

\bibitem{le2019sdr}
Jonathan~Le Roux, Scott Wisdom, Hakan Erdogan, and John~R. Hershey,
\newblock ``{SDR - Half-baked or Well Done?},''
\newblock in {\em {ICASSP}}. 2019, pp. 626--630, {IEEE}.

\bibitem{wichern2019wham}
Gordon Wichern, Joe Antognini, Michael Flynn, Licheng~Richard Zhu, Emmett
  McQuinn, Dwight Crow, Ethan Manilow, and Jonathan~Le Roux,
\newblock ``{WHAM!: Extending Speech Separation to Noisy Environments},''
\newblock in {\em {INTERSPEECH}}. 2019, pp. 1368--1372, {ISCA}.

\bibitem{chung2018voxceleb2}
Joon~Son Chung, Arsha Nagrani, and Andrew Zisserman,
\newblock ``{VoxCeleb2: Deep Speaker Recognition},''
\newblock in {\em {INTERSPEECH}}. 2018, pp. 1086--1090, {ISCA}.

\bibitem{zeghidour2021wavesplit}
Neil Zeghidour and David Grangier,
\newblock ``{Wavesplit: End-to-End Speech Separation by Speaker Clustering},''
\newblock {\em {IEEE ACM Trans. Audio Speech Lang. Process.}}, vol. 29, pp.
  2840--2849, 2021.

\bibitem{YangLW22}
Lei Yang, Wei Liu, and Weiqin Wang,
\newblock ``{TFPSNet: Time-Frequency Domain Path Scanning Network for Speech
  Separation},''
\newblock in {\em {ICASSP}}. 2022, pp. 6842--6846, {IEEE}.

\bibitem{RixenR22}
Joel Rixen and Matthias Renz,
\newblock ``{SFSRNet: Super-resolution for Single-Channel Audio Source
  Separation},''
\newblock in {\em {AAAI}}. 2022, pp. 11220--11228, {AAAI Press}.

\end{thebibliography}

\end{document}